# Spin-Down of Radio Millisecond Pulsars at Genesis

Thomas M. Tauris[1,2*]

[1] Argelander-Institut für Astronomie, Universität Bonn, Auf dem Hügel 71, 53121 Bonn, Germany

[2] Max-Planck-Institut-für Radioastronomie, Auf dem Hügel 69, 53121 Bonn, Germany

[*] To whom correspondence should be addressed. E-mail: tauris@astro.uni-bonn.de

# Spin-Down of Radio Millisecond Pulsars at Genesis

**Millisecond pulsars are old neutron stars that have been spun up to high rotational frequencies via accretion of mass from a binary companion star. An important issue for understanding the physics of the early spin evolution of millisecond pulsars is the impact of the expanding magnetosphere during the terminal stages of the mass-transfer process. Here I report binary stellar evolution calculations that show that the braking torque acting on a neutron star, when the companion star decouples from its Roche-lobe, is able to dissipate >50% of the rotational energy of the pulsar. This effect may explain the apparent difference in observed spin distributions between x-ray and radio millisecond pulsars and help account for the noticeable age discrepancy with their young white dwarf companions.**

Millisecond pulsars (MSPs) are rapidly spinning, strongly magnetized neutron stars. They form as the result of stellar cannibalism, where matter and angular momentum flow from a donor star to an accreting neutron star (*1, 2*). During this process the system is detectable as an x-ray source (*3*). In some cases, x-ray pulsations reveal a fast spinning neutron star (*4*); the 13 known accreting x-ray MSPs (AXMSPs) have an average spin period of $\langle P \rangle_{AXMSP} = 3.3$ ms (*5*). These AXMSPs are thought to be the evolutionary progenitors of radio MSPs. When the donor star decouples from its Roche lobe (*6*) and the mass transfer generating the x-rays ceases, the radio emission is activated (*7*). Until now more than 200 recycled radio MSPs have been detected in our galaxy, both in the field and in globular clusters, with spin periods between 1.4 and 20 ms. These MSPs, which are observed after the Roche-lobe decoupling phase (RLDP), have $\langle P \rangle_{MSP} = 5.5$ ms (*5*) [see the supporting online material, SOM (*8*) for a critical discussion on selection effects and statistics]. It is unknown whether this spin difference is caused by the role of the RLDP or subsequent spin-down from magnetodipole radiation during the lifetime of the radio MSPs.

The interplay between the magnetic field of a neutron star and the conducting plasma of the accreted material is known to provide the accretion torque necessary to spin up the pulsar (*9–11*). However, these interactions can also lead to a torque reversal under certain conditions (*12*), especially when the mass-transfer rate decreases (*13*). Similarly, detailed studies of low-mass x-ray binaries (LMXBs) using stellar evolution codes have demonstrated (*14–17*) that these systems undergo a long, stable phase of mass transfer, which provides sufficient mass to spin up the accreting neutron star to spin periods of milliseconds. However, these previous studies did not combine such numerical stellar evolution calculations with computations of the resulting accretion torque at work. Here, I compute this torque during the termination of the mass-transfer phase by integrating these two methods, using the time-dependent mass-transfer rate to follow its effect on the pulsar spin rate.

The computation of an LMXB donor star detaching from its Roche lobe is shown in Fig. 1 [see further details in the SOM (*8*)]. The full length of the mass-transfer phase was about 1 billion years (Gyr); the RLDP happened in the last 200 million years (Myr) when the mass-transfer rate decreased rapidly. The original mass of the donor star was $1.0\,M_\odot$ and by the time it entered the final stage of the RLDP it had lost 99% of its envelope mass encapsulating a core of $0.24\,M_\odot$ (i.e. the mass of the hot white dwarf being formed after the RLDP). The orbital period at this stage was 5.1 days and the MSP had a mass of $1.53\,M_\odot$. By using the received mass-transfer rate, $\dot{M}(t)$ from my stellar models, the radius of the magnetospheric boundary of the pulsar, located at the inner edge of the accretion disk, can be written as $r_{mag} = \varphi \cdot r_A$ [with $r_A$, the Alfvén radius (*18*), calculated (*8*) following standard prescriptions in the literature (*11, 19*)]. Knowledge of the relative location of $r_{mag}$, the corotation radius, $r_{co}$, and the light cylinder radius, $r_{lc}$, enabled a computation of the accretion torque acting on the pulsar (*8*) (fig. S1). After a long phase of mass transfer as an LMXB the pulsar was spinning at its equilibrium period when entering the RLDP. The rapid torque reversals (unresolved in the center panel of Fig. 1; see also fig. S1) originated from successive, small episodes of spin-up or spin-down depending on the relative location of $r_{co}$ and $r_{mag}$, which reflects the small fluctuations in $\dot{M}(t)$. However, this equilibrium was broken when $\dot{M}(t)$ decreased substantially on a short timescale. That resulted in $r_{mag}$ increasing on a timescale ($t_{RLDP}$) faster than the spin-relaxation timescale, $t_{torque}$, at which the torque would transmit the effect of deceleration to the neutron star and therefore $r_{mag} > r_{co}$. In this propeller phase (*12*), a centrifugal barrier arose and expelled material entering the magnetosphere whereby a braking torque acted to slow down the spin rate of the pulsar even further (see bottom panel in Fig. 1 and fig. S2). The spin-relaxation timescale is given by $t_{torque} \simeq J/N$ which yields:

$$t_{torque} \simeq 50 \, \text{Myr} \cdot B_8^{-8/7} \left( \frac{\dot{M}}{0.1 \dot{M}_{Edd}} \right)^{-3/7} \left( \frac{M}{1.4 M_\odot} \right)^{17/7} \quad (1)$$

where the spin angular momentum of the neutron star is $J = 2\pi I / P$ and the braking torque at the magnetospheric boundary is roughly given by $N \sim \dot{M} \sqrt{GM r_{mag}}$ (*8*). Here $G$ is the gravitational constant; $M$ and $I$ are the neutron star mass and moment of inertia, respectively; and $B_8$ is the surface magnetic flux density in units of $10^8$ G. In addition to this torque, the magnetic field drag on the disk (*20*) was included in the model, although this effect is usually less dominant. In these calculations, I assumed that the strength of the neutron star B-field has reached a constant, residual level before the propeller phase, a reasonable assumption given that less than 1% of the donor star envelope mass remained to be transferred. The propeller phase was terminated when $r_{mag} > r_{lc}$. At this point the MSP activated its radio emission and turned on a plasma wind which then inhibited any further accretion onto the neutron star (*13, 21*). The duration of the RLDP in this example was $t_{RLDP} > 100$ Myr (including early stages before the propeller phase), which is a substantial fraction of the spin-relaxation time scale, $t_{torque} \approx 200$ Myr calculated by using Eq. 1 at the onset of the propeller phase. For this reason the RLDP has an important effect on the spin rate, $P_0$ and the characteristic age of the radio MSP at birth, $\tau_0$. In the case reported here, the radio MSP was born (recycled) with $P_0 \approx 5.4$ ms and an initial so-called characteristic age $\tau_0 \equiv P_0 / 2\dot{P}_0 \approx 15$ Gyr. The spin period before the RLDP was about 3.7 ms, implying that this MSP lost more than 50% of its rotational energy during the RLDP. If the pulsar had not broken its spin equilibrium during the RLDP, it would have been recycled with a slow spin period of $P_0 \approx 58\text{ ms}$ and thus not become an MSP (*22*).

Radio pulsars emit magnetic dipole radiation as well as a plasma wind (*19, 23*). These effects cause rotational energy to be lost, and hence radio MSPs will slow down their spin rates with time after the RLDP. However, they cannot explain the apparent difference in spin distributions between AXMSPs and radio MSPs, because radio MSPs, which have weak surface magnetic field strengths, could not spin down by the required amount even in a Hubble time. The true age of a pulsar (*23*) is given by $t = P/((n-1)\dot{P})\left[1-(P_0/P)^{n-1}\right]$. Assuming an evolution with a braking index $n=3$ and $B=1.0\times10^8\text{ G}$, the time scale $t$ is larger than 10 Gyr, using $P_0 = \langle P \rangle_{AXMSP} = 3.3\text{ ms}$ and $P(t) = \langle P \rangle_{MSP} = 5.5\text{ ms}$. To make things worse, one has to add the main-sequence lifetime of the LMXB donor star, which is typically $3-12\text{ Gyr}$, thereby reaching unrealistic large total ages. Although the statistics of AXMSPs still has its basis in small numbers, and care must be taken for both detection biases (such as eclipsing effects of radio MSPs) and comparison between various subpopulations (*8*), it is evident from both observations and theoretical work that the RLDP effect presented here plays an important role for the spin distribution of MSPs.

The RLDP effect may also help explain a few other puzzles, for example, why characteristic (or spin-down) ages of radio MSPs often largely exceed cooling age determinations of their white dwarf companions (*24*). It has been suggested that standard cooling models of white dwarfs may not be correct (*25–27*), particularly for low-mass helium white dwarfs. These white dwarfs avoid hydrogen shell flashes at early stages and retain thick hydrogen envelopes at the bottom of which residual hydrogen burning can continue for several billion years after their formation, keeping the white dwarfs relatively hot (~$10^4$ K) and thereby appearing much younger than they actually are. However, it is well known that the characteristic age is not a trustworthy measure of true age (*28*), and the RLDP effect exacerbates this discrepancy even further. In the model calculation presented in Fig. 1, it was assumed that $B = 1.0 \times 10^8$ G and $\varphi = 1.0$. However, $P_0$ and $\tau_0$ depend strongly on both $B$ and $\varphi$. This is shown in Fig. 2 where I have calculated the RLDP effect for different choices of $B$ and $\varphi$ by using the same stellar donor model [i.e., same $\dot{M}(t)$ profile] as before. The use of other LMXB donor star masses, metallicities, and initial orbital periods would lead to other $\dot{M}(t)$ profiles (*16, 17*) and hence different evolutionary tracks. The conclusion is that recycled MSPs can basically be born with any characteristic age. Thus we are left with the cooling age of the white dwarf companion as the sole reliable, although still not accurate, measure as an age indicator.

A final puzzle is why no sub-millisecond pulsars have been found among the 216 radio MSPs detected in total so far. Although modern observational techniques are sensitive enough to pick up sub-millisecond radio pulsations, the fastest spinning known radio MSP, J1748−2446ad (*29*), has a spin frequency of only 716 Hz, corresponding to a spin period of 1.4 ms. This spin rate is far from the expected minimum equilibrium spin period (*8*) and the physical mass shedding limit of about 1500 Hz. It has been suggested that gravitational wave radiation during the accretion phase halts the spin period above a certain level (*30, 31*). The RLDP effect presented here is a promising candidate for an alternative mechanism, in case a sub-millisecond AXMSP is detected (*8*).

**Acknowledgments:** I gratefully thank N. Langer and M. Kramer for discussions and funding and without whom these results would not be possible and R. Eatough for helpful comments on the SOM. This work was partly supported by the Cluster of Excellence proposal EXC 1076, "The Nature of Forces and Matter", at the University of Bonn. Radio pulsar data has been obtained from the *ATNF Pulsar Catalogue* (http://www.atnf.csiro.au/research/pulsar/psrcat/).

**Supporting Online Material**

Materials and Methods

SOM Text

Figs. S1 and S2

References (*38–54*)

**Figure 1: Final stages of mass transfer in an LMXB**. The gradual decoupling of the donor star from its Roche lobe causes the received mass-transfer rate, $\dot{M}$ (i.e. the ram pressure of the inflowing material) to decrease whereby the magnetospheric boundary of the neutron star, $r_{mag}$, moves outward relative to the corotation radius, $r_{co}$, and the light cylinder radius, $r_{lc}$ (top). The alternating spin-up/spin-down torques during the equilibrium spin phase are replaced by a continuous spin-down torque in the propeller phase ($r_{co} < r_{mag} < r_{lc}$), until the pulsar activates its radio emission and the magnetodipole radiation remains as the sole braking mechanism, [center; see also (*8*) and figs. S1 and S2]. After an initial phase of spin-down, the spin equilibrium is broken which limits the loss of rotational energy of the pulsar and sets the value of the spin period, $P_0$, of the radio MSP at birth (bottom). Also shown in the bottom panel is the resulting characteristic age, $\tau_0 \propto P_0^2$ of the recycled radio MSP, assuming a constant B-field of $10^8$ G during the RLDP. The gray shaded regions indicate the propeller phase.

**Figure 2: Evolutionary tracks during the Roche-lobe decoupling phase (RLDP)**. Computed tracks are shown as arrows in the $P\dot{P}$–diagram calculated by using different values of the neutron star B-field strength. The various types of arrows correspond to different values of the magnetospheric coupling parameter, $\varphi$. The gray shaded area indicates all possible birth locations of recycled MSPs calculated from one donor star model. The solid lines represent characteristic ages, $\tau$, and the dotted lines are spin-up lines calculated for a magnetic inclination angle, $\alpha = 90°$. The star indicates the radio MSP birth location for the case presented in Fig. 1. The two triangles indicate approximate locations (*36*) of the AXMSPs SWIFT J1756.9–2508 (upper) and SAX 1808.4–3658 (lower). Observed MSPs in the Galactic field are shown as dots [data taken from the *ATNF Pulsar Catalogue*, December 2011]. All the measured $\dot{P}$ values are corrected for the Shklovskii effect, a kinematic projection effect that affects the apparent value of $\dot{P}$ for pulsars (*37*). If the transverse velocity of a given pulsar was unknown, I used a value of 72 km s$^{-1}$, the median value of the 44 measured MSP velocities. The average spin periods of AXMSPs and radio MSPs are indicated with arrows at the bottom of the diagram (*8*).

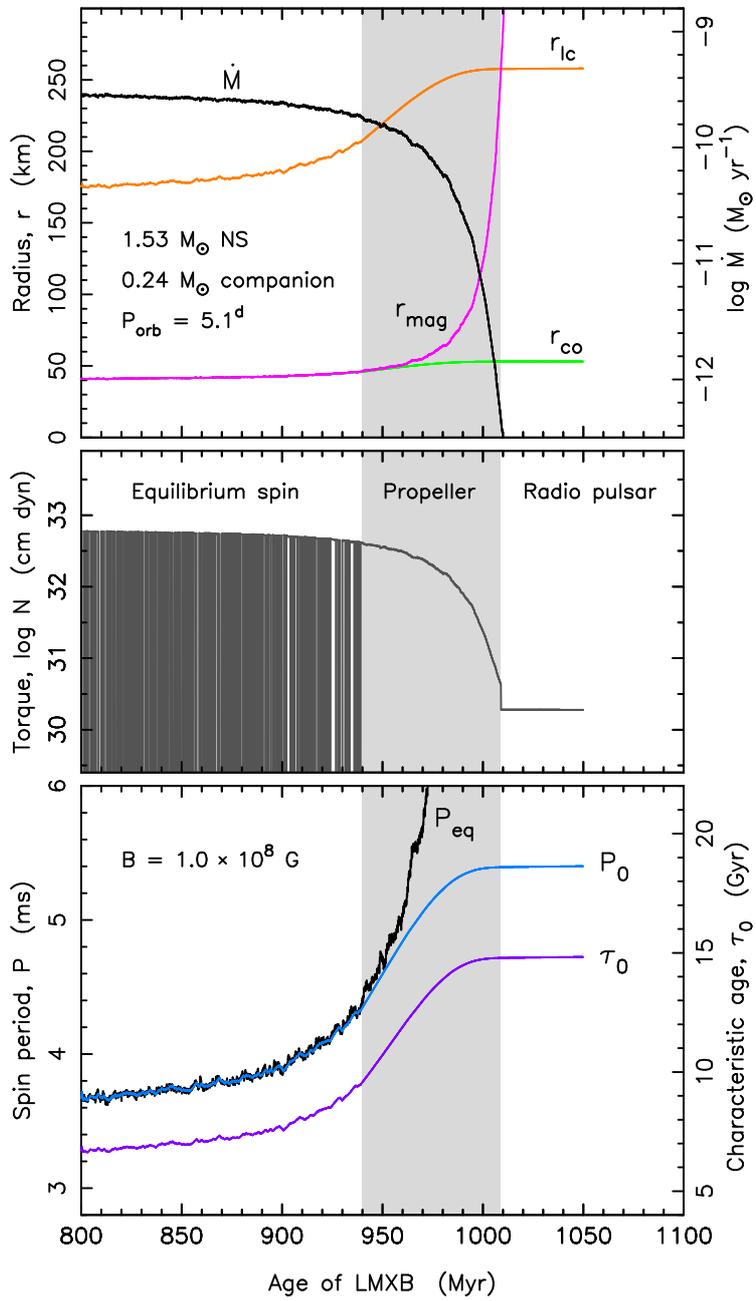

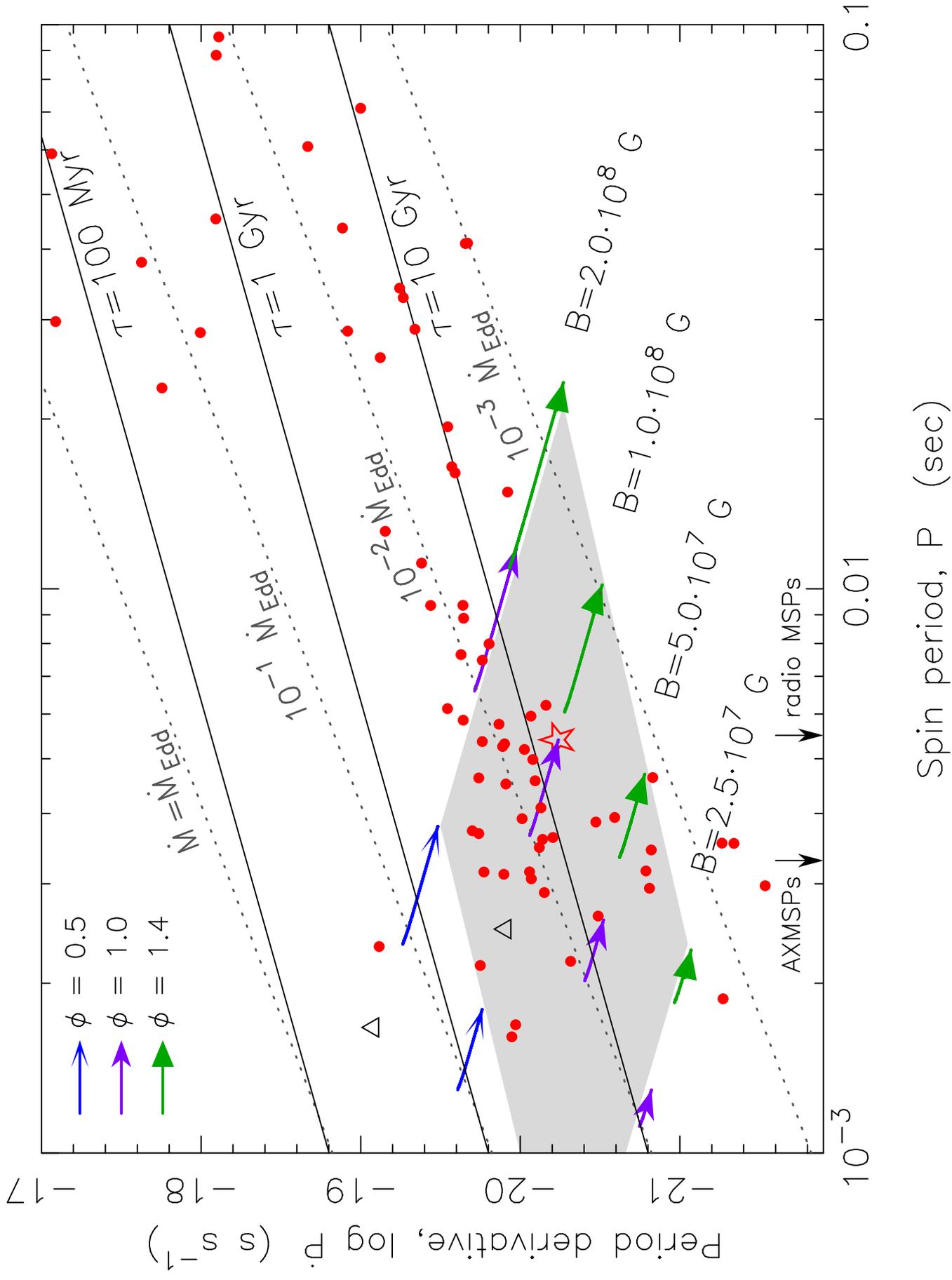

Supporting Online Material

**Numerical binary stellar evolution code**

When modeling the evolution of an X-ray binary one should take into account a number of issues related to stellar evolution, for example, the stability of the mass-transfer process, possible spin-orbit couplings, accretion of material onto the neutron star and ejection of matter from the system. The major uncertainties are related to the specific orbital angular momentum of ejected matter − and for close systems also the treatment of spin-orbit couplings. I applied the Langer code (*38, 39*) which is a binary stellar evolution code developed on the basis of a single star code (*40*) and adapted to binary interactions. It is a 1-dimensional implicit Lagrangian code which solves the hydrodynamic form of the stellar structure and evolution equations (*41*). The evolution of two stars and, in case of mass transfer, the evolution of the mass-transfer rate and of the orbital separation are computed simultaneously through an implicit coupling scheme using standard prescriptions for the Roche-approximation (*42*) and computation of the mass-transfer rate (*43*). The stellar models are computed using OPAL opacities and extended nuclear networks. An updated version of the code has been applied to low- and intermediate-mass X-ray binaries (*44*).

In close orbit LMXBs the mass-transfer rate is sub-Eddington at all times. To account for the relatively small masses of some radio pulsars descending from such systems (*16*), I assumed that the received mass-transfer rate, $\dot{M} = 0.30\,\dot{M}_2$, where $\dot{M}_2$ is the RLO mass-transfer rate from the donor star. Assuming fully conservative RLO does not diminish the RLDP effect discussed here. In the presented model I assumed an initial neutron star mass of 1.30 $M_\odot$ and an orbital period of 0.82 days at the onset of the LMXB phase.

**Magnetospheric boundary and resultant accretion torque**

The light cylinder radius is defined by: $r_{lc} \equiv c/\Omega$, where $\Omega = 2\pi/P$ and $c$ is the speed of light in vacuum. This radius marks the outer boundary of closed B-field lines (perpendicular to the spin axis) of the neutron star. The Alfvén radius is roughly a measure of the boundary of the pulsar magnetosphere, inside of which the flow pattern of accreted plasma is largely dictated by the B-field corotating with the neutron star. It is found by equating the magnetic energy density ($B^2/8\pi$) to the ram pressure of the spherically accreted matter and is approximately given by (*45*): $r_A = B^{4/7} R^{12/7} \left(\dot{M}\sqrt{2GM}\right)^{-2/7}$, where $R$ and $M$ are the radius and mass of the neutron star, respectively, and $G$ is the gravitational constant. The plasma flow onto the neutron star is not spherical, but instead forms an accretion disk where excess angular momentum is transported outwards by turbulent-enhanced viscous stresses (*10, 11, 19, 46*). The location of the inner edge of the disk, i.e. the coupling radius of the magnetosphere, is then estimated as: $r_{mag} = \varphi \cdot r_A$, where $0.5 < \varphi < 1.4$ (cf. refs. *10, 11, 32–34*). The corotation radius, $r_{co} = \left(GM/\Omega^2\right)^{1/3}$ is defined as the radial distance at which the Keplerian angular frequency is equal to the angular spin frequency of the neutron star. Using binary stellar models a mass-transfer profile, $\dot{M}(t)$ is obtained, and which is needed, in combination with the relative locations of $r_{co}$, $r_{mag}$ and $r_{lc}$, to calculate the resulting accretion torque as a function of time, given by:

$$N(t) = n(\omega)\left[\dot{M}(t)\sqrt{GM\,r_{mag}(t)}\,\xi + \frac{\mu^2}{9\,r_{mag}^3(t)}\right] - \frac{\dot{E}_{dipole}(t)}{\Omega(t)} \qquad (S1)$$

Here the second term in the parenthesis on the right-hand side is the drag from the accretion disk interaction with the B-field of the neutron star (20) and $\dot{E}_{dipole} = (-2/3c^3)|\ddot{\mu}|^2$ is the loss of rotational energy due to emission of magnetodipole waves. Following previous work (10, 34, 47) a dimensionless quantity, here defined by: $n(\omega) = \tanh\left(\dfrac{1-\omega}{\delta\omega}\right)$, is introduced to model a gradual torque change in a transition zone near the magnetospheric boundary. It is typically a good assumption that the width of this zone is small (47), i.e. $\delta\omega \ll 1$, corresponding to a step function $n(\omega) = \pm 1$ (Fig. S1).

In case the X-ray pulsar is a fast rotator ($\Omega_{PSR} > \Omega_K(r_{mag})$, yielding a fastness parameter $\omega > 1$, which corresponds to $r_{mag} > r_{co}$) a centrifugal barrier arises causing the magnetosphere to act like a propeller. The acceleration of the incoming gas to escape speed is accomplished by the "magnetic slingshot" mechanism (48), causing the rotational energy of the neutron star to be lost. It has been noted (20, 34, 47), however, that for accretion at a fastness parameter near 1, the velocity excess of neutron star rotation over the Keplerian velocity may not be energetically sufficient to gravitationally unbind all accreting matter, and therefore the expelled gas will return to the disk and build up in mass until the magnetosphere is pushed inward and forces accretion. However, it is important to note that even if the accretion flow is not continuous in nature, in a relatively narrow interval of accretion near a fastness parameter of 1, this suggested cyclic behavior disappears shortly after the onset of the propeller phase when the magnetosphere moves outward rapidly. Furthermore, recent work by (*34*) shows that cyclic accretion is supposed to confine a considerable amount of mass to the disk and even increase the braking torque, exacerbating the loss of rotational energy of the pulsar during the Roche-lobe decoupling phase (RLDP).

**Broken spin equilibrium**

The small fluctuations in $\dot{M}(t)$ are reflected in the location of $r_{mag}$ which then causes rapid oscillations in the sign of the accretion torque, i.e. alternate episodes of spin-up or spin-down. This causes the pulsar spin period to fluctuate about the average value of $P_{eq}$ at any time as already predicted in earlier work (*49*). This behavior is demonstrated in Fig. S2 (which is a zoom-in of the bottom panel of Fig. 1 in the main paper). It is therefore expected that among the population of AXMSPs some systems will be detected with positive (spin-up) torques while others will have a negative (braking) torque acting. The torque reversal timescale is difficult to calculate accurately because of both numerical computational issues and unknown details of the accretion disk physics (*11, 46*). The time steps during the calculated RLDP were less than $10^5$ yr and thus much smaller than $t_{RLDP}$ (100 Myr). However, it should be noted that the torque reversals described here, which are related to fluctuations in $\dot{M}(t)$, appear on a much longer timescale compared to the frequent torque reversals observed in many X-ray binaries (*3*). These torque reversals occur on a timescale of a few years, or decades, and are possibly related to warped disks, disk-magnetosphere instabilities, X-ray irradiation effects of the donor star or transitions in the nature of the disk flow.

From Fig. S2 it is clear that the spin of the neutron star is able to remain in equilibrium during the initial stages of the RLDP. However, the spin equilibrium is broken at some point when $\dot{M}(t)$ decreases too quickly for the torque to transmit the deceleration to the spinning neutron star (see arrow). From this point onwards, $r_{co}$ cannot keep up with the rapidly increasing value of $r_{mag}$. The result of the computation presented here is that more than 50% of the rotational energy ($E_{rot} = \tfrac{1}{2} I \Omega^2$) of the pulsar is lost by the time the radio emission turns on. Other $\dot{M}(t)$ profiles would lead to different results, depending on the RLDP timescale relative to the spin-relaxation timescale in any given case.

**The minimum spin period**

The fastest spin a recycled pulsar can obtain is the minimum equilibrium period (*2*):

$$P_{min} \simeq 0.71\,\text{ms} \cdot B_8^{6/7} \left(\frac{\dot{M}}{0.1 \cdot \dot{M}_{Edd}}\right)^{-3/7} \left(\frac{M}{1.4 \cdot M_\odot}\right)^{-5/7} R_{10}^{18/7} \qquad (S2)$$

where $R_{10}$ is the neutron star radius in units of 10 km and $\dot{M}_{Edd}$ is the Eddington accretion limit (of the order $3.0 \times 10^{-8}\,M_\odot\,yr^{-1}$ for spherical accretion, depending on the hydrogen mass fraction of the transferred material). Hence, in principle sub-millisecond pulsars should be able to form. There are, however, selection effects against detecting sub-ms AXMSPs. The reason is that these objects should preferentially be neutron stars with intrinsic weak B-fields (alternatively, their field could be temporarily buried in the crust at this stage (*50*)) and hence there may not be any observable modulation in the X-ray flux. The burst source XTE J1739−285 has been announced (*51*) to have a neutron star spin frequency of 1122 Hz (corresponding to a spin period of 0.89 ms). However, this result is still not confirmed. There are a number of suggestions − mainly related to gravitational wave radiation (*30, 31*) − why the spin rate of accreting neutron stars seems to be halted at a certain frequency cutoff near 730 Hz (cf. recent discussions in refs. *52, 53*). If future observations reveal sub-ms AXMSPs, but no sub-ms radio MSPs, the RLDP effect suggested here could very well be responsible for such a division. Doppler smearing of radio pulsations in tight binary orbits which could cause a selection bias against detection of sub-ms radio pulsars is less serious in present day acceleration searches (at least for dispersion measures, *DM* < 100).

The absence of AXMSPs with high spin frequencies is discussed by (*33*) who make two statements: i) there is no clear evidence for pulsars to have sufficiently low B-fields and large enough mass quadrupoles that gravitational wave emission will limit their spin-up, and ii) the absence of AXMSPs with frequencies > 750 Hz may instead be caused by a combination of their B-field strength and mass-accretion rate, which leads to their magnetospheric boundary (at spin equilibrium) being located at a large distance, thus limiting $P_{eq}$. To settle the question one must perform detailed modeling of, for example, ultracompact X-ray binaries to probe the RLDP timescale relative to the spin-relaxation timescale. Such computations must include assumptions about the equation of state for (semi)degenerate donors (to calculate real time mass-transfer rates), irradiation effects (which are more important in such tight binaries with bloated donors), accretion disk instabilities and magnetic braking laws – a task beyond the scope of this work.

**On the spin distributions and the (sub)populations of AXMSPs and radio MSPs**

When comparing radio MSPs and AXMSPs one should be aware of differences in their binary properties and observational biases. Using data (*ATNF Pulsar Catalogue*) from the 61 binary Galactic disk pulsars with $P < 20$ ms (*5*), I find $\langle P \rangle_{MSP} = 5.42$ ms with a standard deviation, $\sigma=3.88$ ms. For the now 14 known AXMSPs (*36, 54*), I find $\langle P \rangle_{AXMSP} = 3.28$ ms with $\sigma=1.43$ ms. The difference between $\langle P \rangle_{MSP}$ and $\langle P \rangle_{AXMSP}$ is statistically significant at the 99.9% confidence level using a t-test analysis (assuming normally distributed periods). However, when considering the spin distributions, it is important to keep in mind that these two populations are not trivial to compare, for two main reasons: i) observational biases, and ii) selection criteria for sources with different physical properties.

i) As an example, there is an observational bias against detecting the fastest rotating AXMSPs since these pulsars, in general, have the lowest B-fields ($B < 10^8 \, G$) according to the spin-up theory (*2*). As already mentioned, such weak B-field pulsars may not be able to channel the flow of accreted material – a necessary condition for detecting the modulations in the X-ray signal. Similarly, there is a bias against detecting fast spinning radio MSPs in tight binaries since these pulsars are likely to evaporate their companion star in the process and which may result in eclipses of their radio signal (for the two best studied cases PSR B1957+20 and PSR J2051-0827, however, this is only at the 10% level). One could also consider the removal from the AXMSP sample those sources with orbital periods shorter than a few hours since these systems often lead to the formation of isolated MSPs.

ii) Each population can be divided further into subpopulations depending on orbital period, B-field of the pulsar and nature of the companion star (e.g. ultra-light substellar (brown) dwarf, nuclear burning dwarf and, for radio MSPs, helium or carbon-oxygen white dwarf). Similarly, the AXMSPs can be divided into accretion powered and nuclear powered pulsars (persistent and transient burst sources, respectively). If the nuclear powered pulsars are added to the sample of accretion powered AXMSPs (*36*) the result is $\langle P \rangle_{AXMSP} = 2.86$ ms and $\sigma = 1.30$ ms for all 23 AXMSPs. Furthermore, knowledge of the AXMSP B-fields (or spin frequency derivatives) would reveal their location in the $P\dot{P}$−diagram and thus help disentangle the complex evolutionary links between the various subpopulations of AXMSPs and radio MSPs. Unfortunately, until now only two AXMSPs (*36*) have their B-field estimated (see their location in Fig.2). Although many radio MSPs are found with orbital periods up to a few hundred days, so far we have no samples of AXMSPs with orbital periods above one day (which is a puzzle in itself).

In light of the above discussion one could, for example, consider the subpopulation of the 38 Galactic disk binary radio MSPs which have orbital periods less than 30 days, $P<15$ ms, and which do not have carbon-oxygen white dwarf companions, and compare these to the 7 accretion powered AXMSPs with orbital periods above 2 hours (most of which have minimum companion masses $>0.08\,M_\odot$). In this case one finds $\langle P\rangle_{MSP} = 4.11$ ms (σ=2.15 ms) and $\langle P\rangle_{AXMSP} = 2.57$ ms (σ=0.83 ms). The difference between these average spin periods is significant at the 99.7% significance level. In any case, we are indeed still dealing with small number statistics and in particular when taking the subdivisions of the populations into account. Observationally, the RLDP-effect hypothesis presented here can be verified (to become an accepted theory), or falsified, when future surveys discover a significant number of AXMSPs and radio MSPs confined to an interval with similar orbital periods and companion star masses.

SOM references

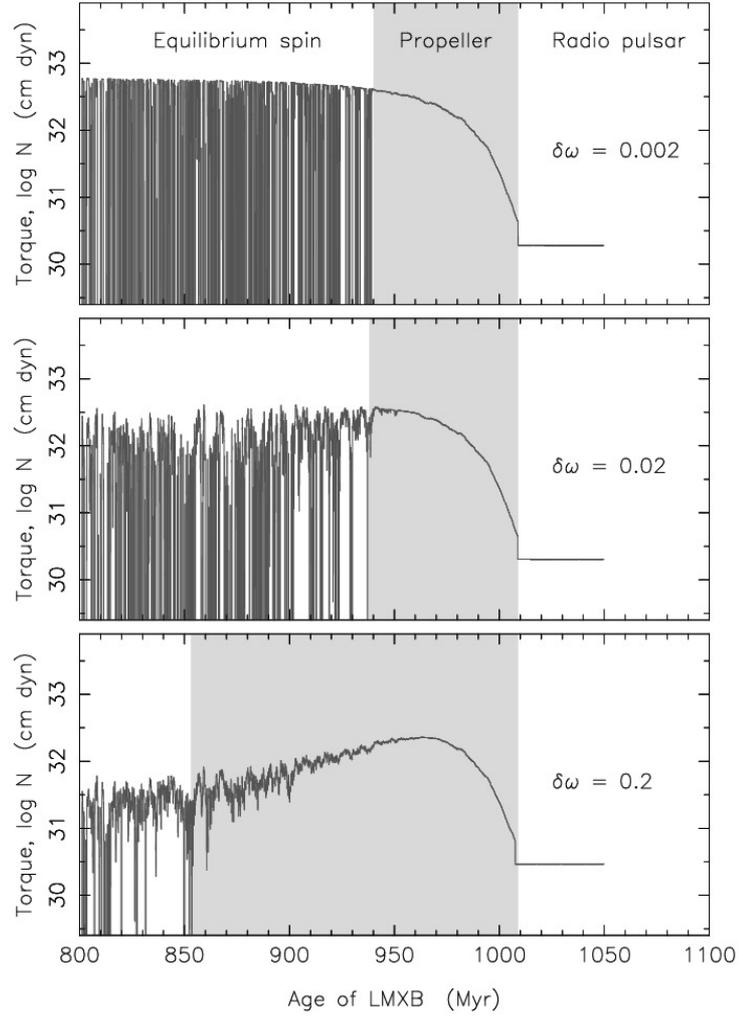

**Figure S1: Braking torque at RLDP.** The resultant braking torque acting on the pulsar during the Roche-lobe decoupling phase is calculated for different values of the transition zone parameter, $\delta\omega$ = 0.002, 0.02 and 0.2 (top to bottom) and $\xi=1$. The post-RLDP spin period, $P_0$ is computed to be: 5.4 ms, 5.3 ms and 4.7 ms, respectively. The pre-RLDP spin period was $P \approx 3.7\;{\rm ms}$. Hence, even at (unrealistic) wide transition zones the RLDP effect is still important. A small value $\delta\omega \ll 1$ corresponds to the switch mode transition $n(\omega) = \pm 1$ used in Fig. 1.

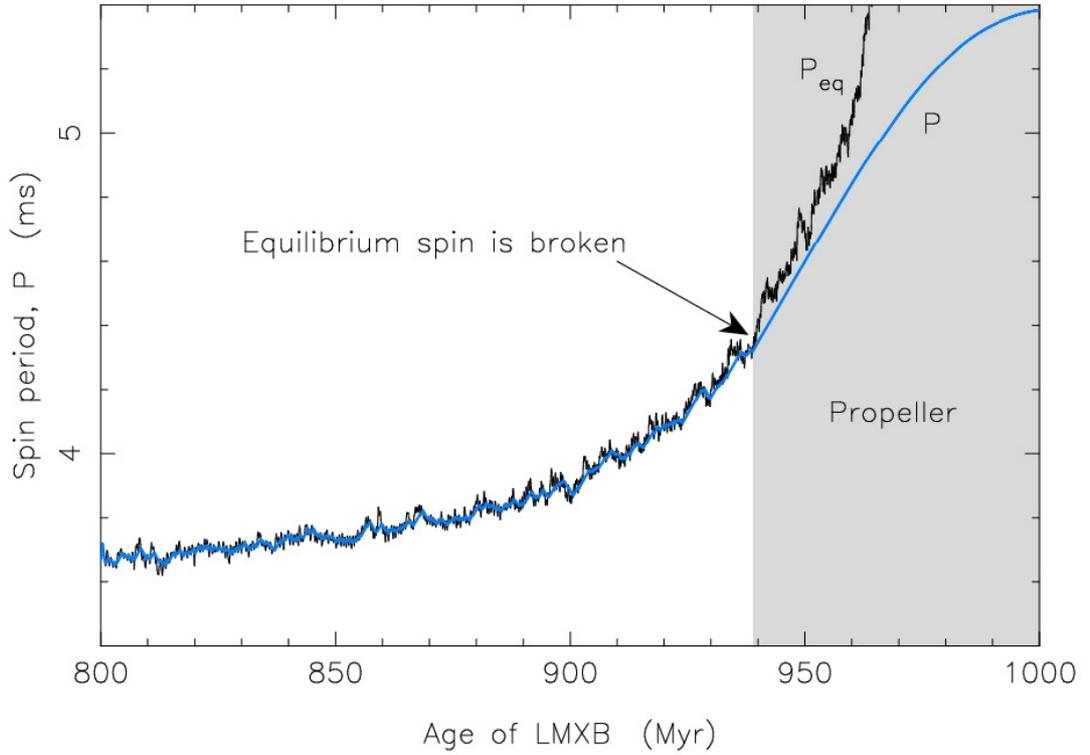

**Figure S2: Transition from equilibrium spin to propeller phase.** At early stages of the Roche-lobe decoupling phase (RLDP) the neutron star spin is able to remain in equilibrium despite the outward moving magnetospheric boundary caused by decreasing ram pressure. However, at a certain point (indicated by the arrow), when the mass-transfer rate decreases rapidly, the torque can no longer transmit the deceleration fast enough for the neutron star to remain in equilibrium. This point marks the onset of the propeller phase where $r_{mag} > r_{co}$ (and $P < P_{eq}$) at all times. This plot is a zoom-in of the bottom panel of Fig. 1 in the main paper.